\documentclass[prl,twocolumn,preprintnumbers]{revtex4-1}%
\usepackage{epsf,epsfig}
\usepackage{amssymb,amsmath,amsfonts}
\usepackage{hyperref}

\usepackage{graphicx}

\def\NN{{\cal N}}

\def\SS{{\cal S}}

\def\L{{\mathfrak L}}
\def\Vol{{\rm Vol}}
\def\p{{\partial}}

\newcommand{\be}{\begin{equation}}
\newcommand{\ee}{\end{equation}}
\newcommand{\beq}{\begin{equation}}
\newcommand{\eeq}{\end{equation}}
\newcommand{\ben}{\begin{displaymath}}
\newcommand{\een}{\end{displaymath}}
\newcommand{\beqa}{\begin{eqnarray}}
\newcommand{\eeqa}{\end{eqnarray}}
\newcommand{\bea}{\begin{eqnarray}}
\newcommand{\eea}{\end{eqnarray}}
\newcommand{\bean}{\begin{eqnarray*}}
\newcommand{\eean}{\end{eqnarray*}}
\newcommand{\ba}{\begin{array}}
\newcommand{\ea}{\end{array}}
\newcommand{\bi}{\begin{itemize}}
\newcommand{\ei}{\end{itemize}}

{}{}
{}{}
\def\vereq#1#2{\lower3pt\vbox{\baselineskip1.5pt \lineskip1.5pt
\ialign{$\m@th#1\hfill##\hfil$\crcr#2\crcr\sim\crcr}}}
\makeatother

\begin{document}

\title{Holographic entanglement entropy in open-closed string duality}

\preprint{DCPT-2017/01}

\author{Vasilis Niarchos}
\affiliation{Department of Mathematical Sciences and Center for Particle Theory\\
Durham University, Durham, DH1 3LE, UK}

\begin{abstract}
\noindent
We study minimal co-dimension-2 surfaces in the asymptotically flat background of extremal 3-brane 
solutions in ten-dimensional type IIB supergravity. A conjectured
open-closed string duality combined with the Ryu-Takayanagi prescription implies 
that the area of the surfaces we consider could be interpreted as the entanglement
entropy of a dual (3+1)-dimensional large-$N$, strongly-coupled open string field theory on D3-branes. 
As the size of the surface is varied we observe a transition from a volume law to an area law in agreement
with expectations from non-locality in an open string field theory.
Some of the specifics of this transition bear a qualitative resemblance with the behaviour of holographic entanglement
entropy in non-commutative super-Yang-Mills theory. 
\end{abstract}
\maketitle

\section{Introduction}

A holographic correspondence between theories of gravity in asymptotically flat backgrounds 
and open string theories, generalizing the AdS/CFT correspondence beyond the near-horizon limit 
\cite{Maldacena:1997re}, has been conjectured by several authors in the past, see e.g.\ 
\cite{Douglas:1996yp,Gubser:1998kv,deAlwis:1998mi,Gubser:1998iu,Park:1999xz,Intriligator:1999ai,Khoury:2000hz,Danielsson:2000ze,Park:2001bm,Polyakov:2001af,DiVecchia:2003ae,Amador:2003ju,DiVecchia:2005vm,Niarchos:2015moa,Grignani:2016bpq}. For D3-branes in ten-dimensional type IIB supergravity 
the conjecture proposes a holographic relation between closed string theory in 3-brane solutions
and the large-$N$, strongly-coupled $U(N)$ string field theory on D3-branes. 

More recently in \cite{Niarchos:2015moa} the following points were put forward/emphasized: 
$(i)$ General open-closed string dualities may arise for string field theories on D-brane setups 
(with suitable co-dimension) in generic closed string backgrounds. The asymptotic closed string background
plays the role of an arbitrary external source for open string fields.
$(ii)$ The absence of a near-horizon decoupling limit in these dualities may be justifiable   
as a consequence of the open string completeness conjecture by A.\ Sen \cite{Sen:2003iv,Sen:2004nf}. 
Non-trivial checks of this conjecture were presented in open string tachyon condensation and 
non-critical string theory.
$(iii)$ In this more general version of the holographic principle it is proposed that 
holography works as a {\it tomographic} principle, where different open string field theories capture/reconstruct 
holographically different subsectors of closed string theory/gravity.

Finding evidence in favor of this proposal, and setting up a solid holographic dictionary,
is a highly non-trivial task, largely hindered at the present stage by the lack of powerful
computational tools in interacting open string field theories. This is unlike the
AdS/CFT correspondence where many non-trivial checks can be found by matching explicit 
(super)gravity computations to corresponding computations in strongly-coupled large-$N$ quantum field theories.
With such technical limitations it is useful to identify structures in classical gravity that reproduce familiar features from 
open string theory. In \cite{Niarchos:2015moa} (see also \cite{Niarchos:2014maa}) 
we focused on the long-wavelength dynamics of asymptotically flat 
$p$-brane solutions, and argued that the sector related to the abelian (center-of-mass motion, singleton) dynamics 
of the branes could be identified in gravity within the blackfold approach \cite{Emparan:2009cs,Emparan:2009at}, 
and systematically recast in terms
of the abelian Dirac-Born-Infeld (DBI) effective action. The latter is a characteristic feature of open string theory.
For other recent discussions of the emergence of the DBI action from (super)gravity we refer the reader to 
\cite{Hatefi:2012bp,Grignani:2016bpq,Maxfield:2016vpw}.

In this note we are looking for a quantity that has the power to probe more efficiently the 
full non-abelian sector of the dual $U(N)$ string field theory. For concreteness, we will focus
on the canonical example of D3-branes in flat space. The quantity we would like to consider 
is entanglement entropy (EE).
This is an observable that can probe interactions and degrees of freedom 
across different scales. It has been proposed to have a simple holographic manifestation in gravity 
as the area of a corresponding minimal co-dimension-2 surface \cite{Ryu:2006bv,Ryu:2006ef}. 
Although this proposal is better understood in the case of AdS space 
\cite{Casini:2011kv,Hartman:2013mia,Faulkner:2013yia,Lewkowycz:2013nqa}, 
here we will assume that the prescription is more generally valid and applies also to asymptotically flat spacetimes.  
Our main goal is to examine the properties of 
holographic entanglement entropy (HEE) in asympotically flat 3-brane solutions where open-closed string duality
could operate according to the above-mentioned conjectures.

A clear feature of open string theory we should be looking for is non-locality. This is manifest in EE as 
a volume law, rather than an area law, which is characteristic of local quantum field theory 
\cite{Barbon:2008ut,Eisert:2008ur,Rabideau:2015via}. 
The volume law in HEE has been noted before in a related context
in a proposal for flat space holography in \cite{Li:2010dr}. 
However, unlike \cite{Li:2010dr} in this note we are not considering HEE solely
in flat space; we are considering it in a solution that interpolates between a near-horizon AdS space and the asymptotic flat space, where we have a more specific conjecture for the anticipated dual non-gravitational theory.
Besides the volume law, which is mainly due to 
flat space (in analogy to \cite{Li:2010dr}), our case also exhibits  
a transition to the more standard area law as the size of the entangling region is varied. 
We argue that the main features of this result are consistent with expectations from an 
interacting open string field theory. 

The volume law has also been noted in the (3+1)-dimensional 
holographic duals of the dipole theory and non-commutative 
super-Yang-Mills (NCSYM) theory \cite{Barbon:2008ut,Fischler:2013gsa,Karczmarek:2013xxa}. 
We will observe that the asymptotically flat 3-brane HEE exhibits similarities with the HEE of the NCSYM theory.

\section{Main result}
We consider the HEE, $\SS_{A_S}$, of a prospective dual $(3+1)$-dimensional open string field theory on a straight belt 
\beq
\label{mainaa}
A_S = \left\{ x^i \Big | x^1 \in \left[-\frac{\ell}{2},\frac{\ell}{2}\right]~, ~ x^2,x^3 \in \left[-\frac{L}{2},\frac{L}{2}\right] \right\}
~.
\eeq
$L$ is an infrared cutoff that is taken eventually to infinity. 
$\ell$ is the tunable width of the belt.
We will also introduce an ultraviolet length cutoff $\varepsilon$ that is kept fixed throughout the computation.
We are mainly interested in the regime 
\beq
\label{mainaaa}
\varepsilon \ll ( 4\pi \lambda)^{\frac{1}{4}} \ell_s ~,
\eeq 
where the asymptotic radial cutoff $r_{\varepsilon}$ in the 
gravitational bulk is outside the near-horizon region ($r_H \ll r_\varepsilon = \frac{r_H^2}{\varepsilon}$ ---see eq.\ \eqref{grab} for the definition of $r_H$). $\ell_s = \sqrt{\alpha'}$ is the string length.
When the cutoff is well inside the near-horizon throat we recover the 
familiar divergent piece of the $\NN=4$ SYM EE, $\SS_{A_S} \simeq \frac{N^2}{2\pi} \frac{L^2}{\varepsilon^2}$, which
is less interesting for our purposes.  

The main result 
of this note is the following prediction for the HEE of the large-$N$ string field theory on D3-branes in the regime
\eqref{mainaaa}
\beq
\label{mainab}
\frac{\SS_{A_S}(\ell)}{L^2} = \Bigg\{
\begin{array}{c}
\frac{N^2 \sqrt{\lambda}}{\sqrt \pi} \frac{\alpha' \ell}{\varepsilon^5} +\ldots~~, ~~\ell \ll \ell_c
\\
~\\
\frac{2N^2 \lambda}{3} \frac{{\alpha'}^2}{\varepsilon^6} +\ldots~~, ~~\ell \gg \ell_c
\end{array} 
~.
\eeq
$\ell_c\sim \sqrt{\lambda} \frac{\alpha'}{\varepsilon}$ 
is a critical length where a transition between different branches of extremal co-dimension-2 surfaces occurs.
It is estimated numerically in eq.\ \eqref{graj} and with an analytic extrapolation in eq.\ \eqref{grao}.
The dots indicate less divergent terms. 

Eq.\ \eqref{mainab} exhibits a transition between a short distance regime 
with a characteristic volume dependence to a long distance regime with the more standard area dependence. We find
evidence that this is a robust feature of the system independent of the choice of the geometry of the entangling region
$A_S$. A similar computation of the HEE for the cylinder
\beq
\label{disc}
A_S= \left\{ x^i \Big | (x^1)^2 + (x^2)^2 = \ell^2~, ~ x^3 \in \left[-\frac{L}{2},\frac{L}{2}\right] \right\}
\eeq
verifies the same type of transition (see eq.\ \eqref{finalac}) consistent with an interpretation based 
on open-closed string duality. The physical content of eq.\ \eqref{mainab} will be discussed further in 
the last section. In the next section we explain how \eqref{mainab} is derived from gravity.
For economy and clarity of the presentation we will focus exclusively on the derivation of the HEE on the 
straight belt geometry \eqref{mainaa}.

\begin{figure*}[t!]
	\centering
	\includegraphics[width=3.4in]{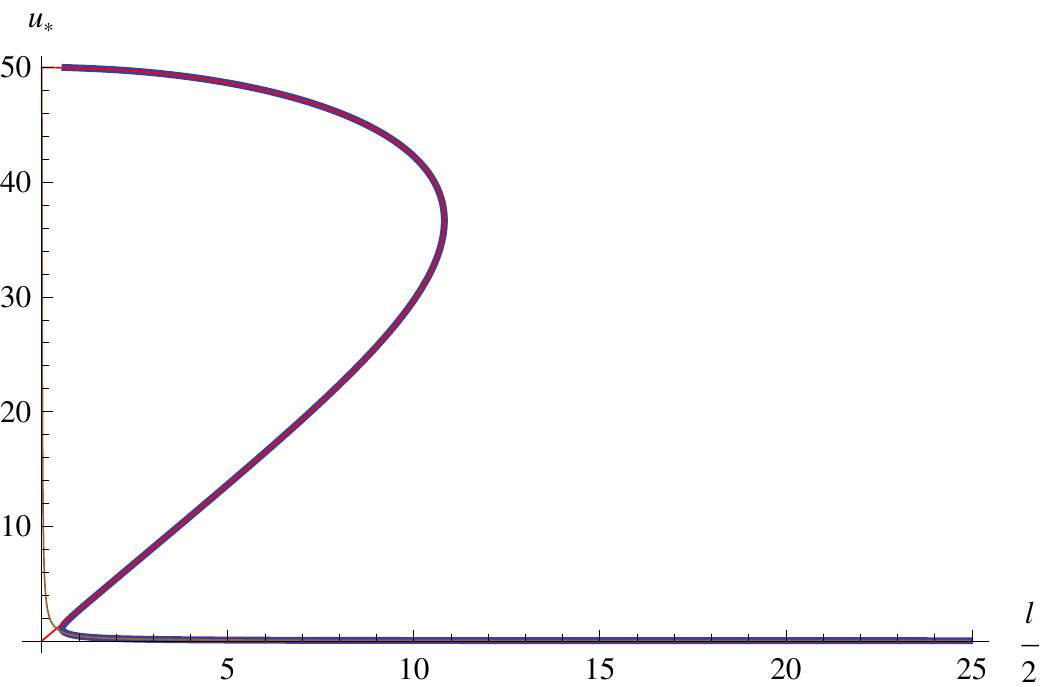}~~~~
	\includegraphics[width=3.4in]{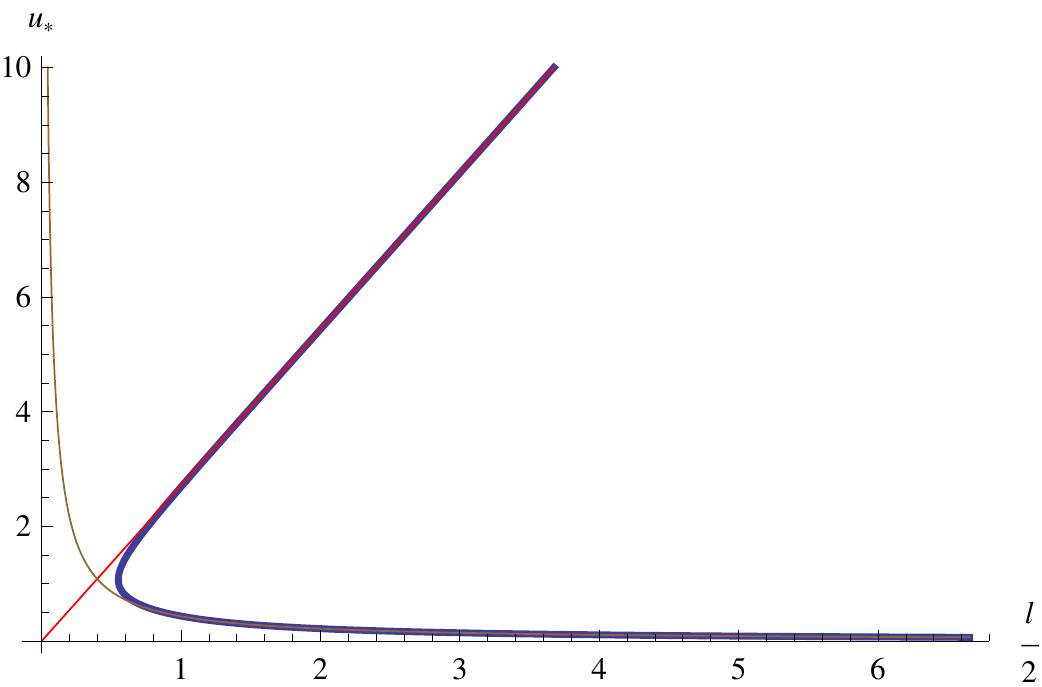}
	\caption[Close up of]
	{A plot of the turning point $u_*$ of the solutions of eq.\ \eqref{graf} as a function of the dimensionless
	belt half-width $\frac{l}{2}$. Both plots represent the solution for the same cutoff value $\epsilon=0.02$ (blue thick 
	line). The right plot zooms around the point where the middle and bottom branches meet. The red curve is an 	
	analytic prediction based on the approximation \eqref{grak}. The brown curve is an analytic prediction based on the 
	approximation \eqref{grbab}.
	}
	\label{ulFigs}
\end{figure*}

\section{HEE from classical gravity}

We focus on $N$ parallel overlapping D3-branes in flat space in ten-dimensional type IIB string theory. In the large-$N$
't Hooft limit, where $\lambda=g_s N$ is kept fixed and large ($g_s$ being the string coupling), 
this setup is most conveniently described by the extremal 3-brane supergravity solution 
\bea
\label{graa}
&&ds^2 = H^{-1/2} \eta_{\mu\nu} dx^\mu dx^\nu + H^{1/2} (dr^2 + r^2 d\Omega_5^2)
~,
\\
&&e^\Phi = g_s~, ~~ C_4 = g_s^{-1} (H^{-1}-1) dx^0\wedge dx^1 \wedge dx^2 \wedge dx^3
~.\nonumber 
\eea
The dilaton $\Phi$ is constant and the 4-form $C_4$ yields a self-dual 5-form flux. 
The solution is expressed here in the string frame. The function $H$ is 
\beq
\label{grab}
H(r) = 1+ \frac{r_H^4}{r^4}~, ~~ r_H^4 = 4\pi \lambda {\alpha'}^2
~.
\eeq
$r\to \infty$ is the asymptotic flat space region.

In what follows we will express all length scales of the problem in terms of the length scale $r_H$ that is intrinsic
to the solution \eqref{graa}. In particular, for the quantities in \eqref{mainaa} 
we set $r= u \, r_H$, and $\ell=l \, r_H$, $L=\L\, r_H$, 
$\varepsilon= \epsilon \, r_H$, where $u,l,\L$ and $\epsilon$ are now dimensionless quantities.

Following the Ryu-Takayanagi prescription \cite{Ryu:2006bv,Nishioka:2006gr,Klebanov:2007ws} 
we are looking for a constant time, co-dimension-2 minimal surface $\bar A_S$
in the background \eqref{graa} that asymptotes (at large radial distance $u$) to the boundary of the region \eqref{mainaa}. The volume of this minimal surface,
\beq
\label{grac}
\Vol(\bar A_S) = \int d^8 \sigma\, e^{-2\Phi} \sqrt{G^{(8)}_{ind}}
~,
\eeq 
can be used to express the HEE as
\beq
\label{grad}
\SS_{A_S} = \frac{\Vol(\bar A_S)}{32\pi^6 (\alpha')^4}
~.
\eeq

In eq.\ \eqref{grac} $G_{ind}^{(8)}$ is the determinant of the induced metric on the co-dimension-2 surface $\bar A_S$.
Wrapping this surface around the 5-sphere of the background \eqref{graa} and the directions $x^2,x^3$, 
and assuming a single dependence on the coordinate $x\equiv \frac{x^1}{r_H}$, we can express it as a solution 
$u(x)$ that extremizes the volume \eqref{grac}, written more explicitly as
\beq
\label{grae}
\Vol(\bar A_S) = \frac{\pi^3 r_H^8 \L^2}{g_s^2} \int dx\, u^5 H^{1/2} \sqrt{1+ H \left( \dot u \right)^2}
~,
\eeq
where $H=1+ u^{-4}$ and $\dot u =\frac{\p u}{\p x}$. 
The extremization equations of \eqref{grae} can be recast as the conservation equation
\beq
\label{graf}
u^5 \sqrt{\frac{1+u^{-4}}{1+\left( 1+ u^{-4} \right) \dot u^2}} = u_*^5 \sqrt{1+u_*^{-4}} \equiv c
~,\eeq
where $u_*$ is the turning point of the surface in the bulk. $c$ is essentially the value of the conserved Hamiltonian
of the system \eqref{grae}.
We solve the non-linear first-order differential equation \eqref{graf} with an explicit UV cutoff $\epsilon$
\beq
\label{grag}
u\left( x=\pm \frac{l}{2}\right) = \frac{1}{\epsilon}
~.\eeq
As was noted in previous work \cite{Karczmarek:2013xxa}, 
it is important to keep the cutoff fixed thoughout the computation, otherwise important
branches of the solution can be missed as $\epsilon \to 0$.

After a few trivial algebraic manipulations, the HEE \eqref{grad} evaluated on the on-shell profile of the function $u(x)$ can be written as
\bea
\label{grai}
\SS_{A_S}= \frac{\L^2 N^2}{\pi c} \int_{u_*}^{\frac{1}{\epsilon}} du \, 
u^{10} \sqrt{
\frac{\left( 1+ u^{-4} \right)^{3}}
{ u^{10} c^{-2} (1+u^{-4}) -1}
}
~.
\eea

The equations \eqref{graf}, \eqref{grag} are easily solved numerically. The turning point $u_*$ as a function of the belt 
width $l$ is plotted in 
Fig.\ \ref{ulFigs} for the cutoff value $\frac{1}{\epsilon}=50$, or $\epsilon = 0.02$ (other small values
of $\epsilon$ were checked to exhibit the same behavior). 
We observe an intermediate range of $l$, where there are three separate 
branches of extermal surface solutions, let us call them upper, middle and bottom branches for decreasing values of 
$u_*$. In the upper branch the extremal surface is mainly embedded in the asymptotic flat space region. In the lower 
branch the extremal surface is well embedded inside the AdS throat. We will analyse each of these branches 
analytically with perturbative methods in a moment. 

\begin{figure*}[t]
	\centering
	\includegraphics[width=3.4in]{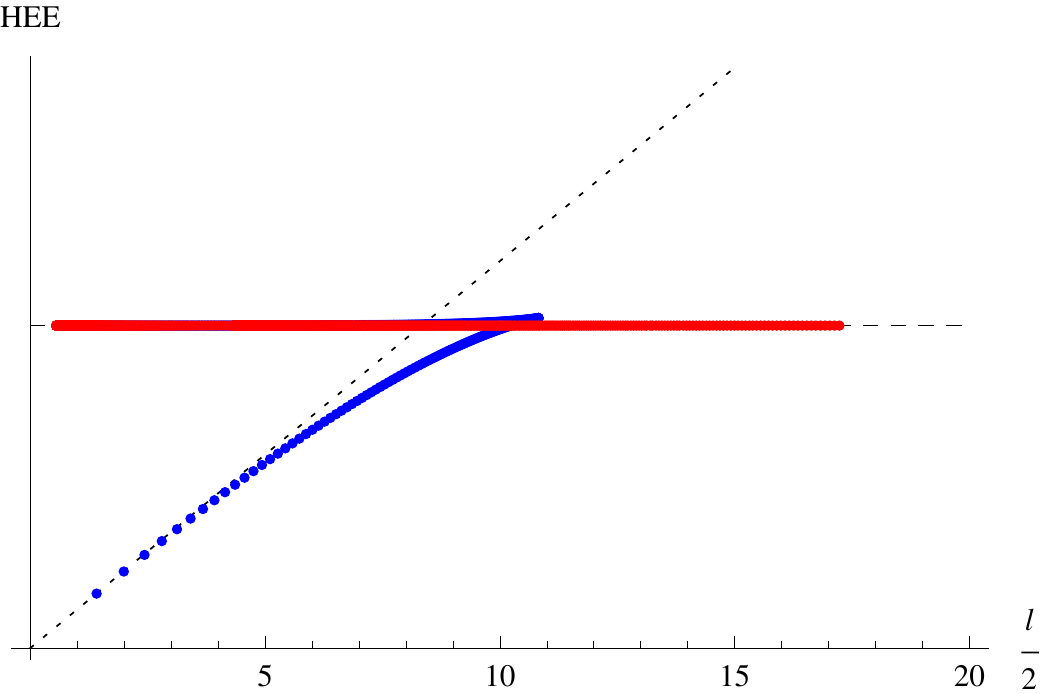}~~~~
	\includegraphics[width=3.4in]{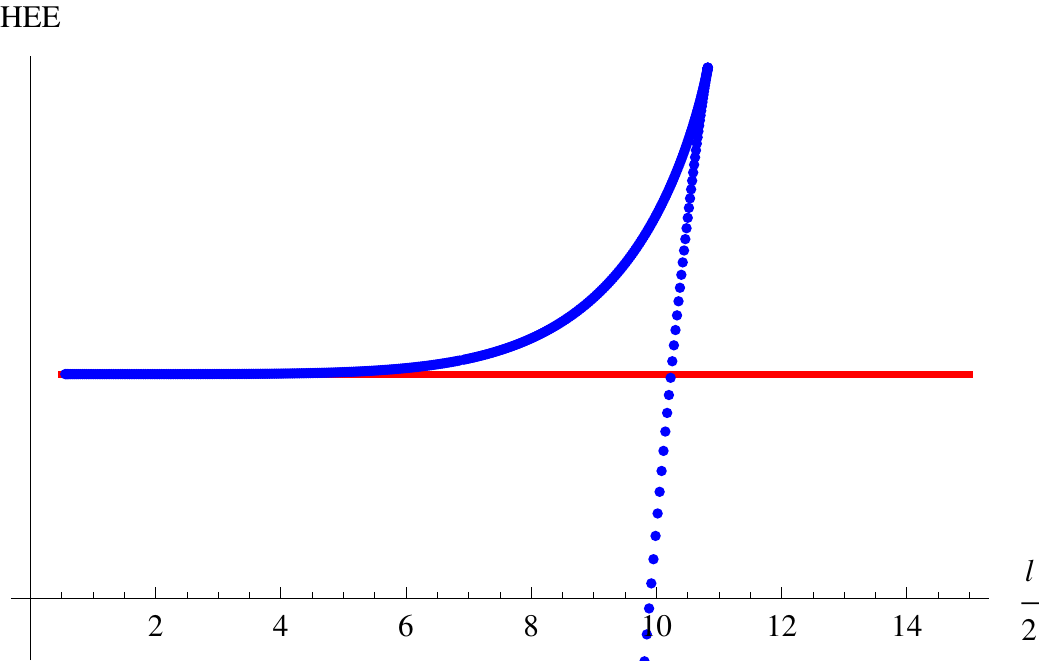}
	\caption[Close up of]
	{A plot of the profile of the HEE as a function of the belt half-width $\frac{l}{2}$. The same cutoff value 
	$\epsilon=0.02$ is used in this figure as in Fig.\ \ref{ulFigs}. The right plot zooms into the region of the transition at 		$l=l_c$ of the left plot. The dotted line that passes through the origin in the left plot depicts the volume law 			\eqref{gram}, which is linear in $l$. The dashed horizontal line depicts the area law \eqref{gran}. The blue points 		refer to the numerical data for the upper and middles branches, and the red points to the data for the bottom branch. 
	}
	\label{HEEFigs}
\end{figure*}

In Fig.\ \ref{HEEFigs} we plot the value of the HEE for each of 
these branches. We observe that the upper branch is the dominant (or single) branch for all values of $l$ in the 
interval $[0,l_c)$. For $l>l_c$ the bottom (AdS) branch takes over and dominates. At very large values of $l$ this 
branch is the only existing branch. 
The existence of a similar pattern of three co-existing extremal surfaces was also observed in 
the NCSYM theory in \cite{Karczmarek:2013xxa}.

Numerically, we observe that the critical width is
\beq
\label{graj}
l_c \sim \frac{0.4}{\epsilon} ~~\Leftrightarrow~~ 
\ell_c \sim \frac{0.4\, r_H^2}{\varepsilon}
=0.8 \sqrt{\pi \lambda} \, \frac{\ell_s}{\varepsilon}\, \ell_s
~.
\eeq
The explicit presence of the UV cutoff scale $\varepsilon$
in $\ell_c$ was also noted in the case of the NCSYM theory \cite{Barbon:2008ut,Fischler:2013gsa,Karczmarek:2013xxa}. 
An analytic rough estimate of $\ell_c$ will 
be discussed soon.

We proceed to analyse each of these branches analytically with approximations performed in two opposite 
regimes: $u_*\sim \frac{1}{\epsilon} \gg 1$ and $u_*\ll \frac{1}{\epsilon}$.

\vspace{0.3cm}
\noindent
{\bf Approximations in the flat space regime: $u_* \gg 1$.} 
In this regime the function $H(u)$ is 1 to first approximation and 
$c\sim u_*^5$. Then, we can easily find an analytic solution for $u(x)$ in the form
\beq
\label{grak}
x = - \frac{\sqrt{\pi}  \Gamma\left( \frac{2}{5} \right) u_*}{\Gamma\left( -\frac{1}{10} \right)}
- \frac{u_*^5}{4u^4} ~_2F_1\left( \frac{2}{5},\frac{1}{2},\frac{7}{5},\frac{u_*^{10}}{u^{10}}\right)
~.\eeq
The relation between $l$ and $\frac{1}{\epsilon}$ follows immediately from this equation 
by implementing the boundary condition \eqref{grag}. We notice
that the resulting curve (red curve in both plots in Fig.\ \ref{ulFigs}) agrees very well with the numerical 
solution in both the upper and middle branches and starts deviating only when $u_*\sim 1$ and $l\sim 1$.

Under the same assumptions, for the HEE we obtain the expression
\bea
\label{gral}
\SS_{A_S} = \frac{\L^2 N^2}{\pi} \frac{u_*^6}{30} 
\bigg[&& \frac{3\sqrt \pi \Gamma\left( -\frac{3}{5} \right)}{\Gamma\left( -\frac{1}{10}\right)}
\\
&&+ 5 \left( \frac{1}{\epsilon u_*}\right)^6 ~_2F_1 \left( -\frac{3}{5},\frac{1}{2},\frac{2}{5},(\epsilon u_*)^{10} \right)\bigg]
~.\nonumber
\eea
It is not hard to show by a series expansion of this equation that when $u_* \sim \frac{1}{\epsilon}$ to leading order
\beq
\label{gram}
\SS_{A_S}  \simeq \frac{N^2}{2\pi} \frac{\L^2 l}{\epsilon^5}
~,\eeq 
which is proportional to the volume of the belt $A_S$, and represents the leading contribution to the HEE when 
$\ell \ll \ell_c$. This result provides the upper line term on the r.h.s.\ of eq.\ \eqref{mainab}). 
The linear expression in $l$, \eqref{gram}, is depicted in the left plot of Fig.\ \ref{HEEFigs} by the dotted line 
that goes through the origin.

In a different limit, where $u_*\ll \frac{1}{\epsilon}$, we can deduce from the series expansion of the expression 
\eqref{gral} the leading HEE contribution
\beq
\label{gran}
\SS_{A_S} \simeq \frac{N^2}{6\pi}\frac{\L^2}{\epsilon^6}
~.
\eeq 
Strictly speaking this result is valid only under the additional assumption $u_*\gg 1$ along the subdominant middle
branch. It is clear, however, from the numerical data depicted in Figs.\ \ref{ulFigs}, \ref{HEEFigs} that the solution
\eqref{grak} works impressively well even for order 1 values of $u_*$. It is also clear from the data of Fig.\ 
\ref{HEEFigs} that once the middle branch asymptotes to the bottom branch, it reaches the value \eqref{gran} 
which remains constant as $l$ is increased. This constant is the same as the leading divergence of the bottom branch 
(see red curve in the plots of Fig.\ \ref{HEEFigs}). 

This observation allows us to obtain a rough estimate of $l_c$ by equating the HEEs in eqs.\ \eqref{gram} 
and \eqref{gran}
\beq
\label{grao}
 \frac{N^2}{2\pi} \frac{\L^2 l_c}{\epsilon^5} \sim \frac{N^2}{6\pi}\frac{\L^2}{\epsilon^6}
 ~~  \Leftrightarrow ~~ l_c \sim \frac{1}{3\epsilon}
 ~.
 \eeq
 Comparing with \eqref{graj} we observe that the factors $\sim 0.4$ (from the numerics) and 
 $\frac{1}{3}\sim 0.3$ (from the analytic approximation), compare well with each other.

\vspace{0.3cm}
\noindent
{\bf Information from the AdS regime: $u_* \ll 1$.} Assume for a moment that we take the radial cutoff 
$r_\varepsilon$ inside the near-horizon region, i.e.\ we take $r_\varepsilon \ll r_H$.  
In this regime we recover the minimal surface as a deformation of the more familiar $AdS_5 \times S^5$ results.
The function $H(u)$ is to first approximation $u^{-4}$ and $c\sim u_*^3$. The analytic solution for the function
$u(x)$ in this case implies at the cutoff
\beq
\label{grbab}
\frac{\ell}{2} = \frac{1}{u_*} \left[ \frac{\sqrt \pi \Gamma \left( \frac{2}{3} \right)}{\Gamma \left( \frac{1}{6} \right)}
- \left( \epsilon u_* \right)^6 ~_2F_1 \left( \frac{1}{2},\frac{2}{3}, \frac{5}{3}, \left( \epsilon u_*\right)^6 \right) \right]
~.\eeq
One can see from the numerical solutions of the minimal surface in the opposite regime $r_H\ll r_\varepsilon$ (i.e.\
the regime \eqref{mainaaa} of interest in this note) 
that the position $x(u)$ of the surface moves very slowly as $u$ decreases until the surface 
reaches the near-horizon throat. 
As a result, the equation \eqref{grbab} is not only a good approximation in the 
AdS regime, $r_\varepsilon \ll r_H$, but also in the regime \eqref{mainaaa}. 
This result is clearly visible in the numerical results of Fig.\ \ref{ulFigs}, where the brown curve 
depicting \eqref{grbab} is in excellent agreement with the bottom branch from the region $u\sim 1$, $l\sim 1$ and
towards higher values of $l$.

\section{Interpretation of results}
The above-mentioned conjectures of open-closed string duality imply, if correct, 
that the HEE \eqref{mainab} is holographically related to the EE
of a dual large-$N$ open string field theory. We cannot verify \eqref{mainab} directly in a strongly interacting 
open string field theory, but it is still natural to ask if a transition like \eqref{mainab} could be envisaged in an open string
theory. Here we would like to argue that the answer to this question is naturally a positive one.

Let us start with the estimate of the transition scale $\ell_c\sim \sqrt \lambda \frac{\alpha'}{\varepsilon}$. The 
$\sqrt \lambda$ factor could clearly be an effect of the strong 't Hooft coupling limit. In a weakly coupled string
theory one would anticipate a transition scale of the form $\ell_c \sim \frac{\alpha'}{\varepsilon}$. Could this be 
consistent? From previous discussions of EE in non-local theories (see e.g.\ \cite{Barbon:2008ut}) 
we know that $\ell_c$ is naturally associated
with the scale of non-locality. In the case of an open string theory the characteristic size of a rotating string with 
energy $E$ is $\ell_E \sim E \alpha'$ \cite{Zwiebach:2004tj}. 
Consequently, in the presence of a UV cutoff $\varepsilon^{-1}$ the 
maximum value of $\ell_E$, which should be associated with the non-locality scale, is 
$\ell_{\varepsilon^{-1}}=\ell_c \sim \frac{\alpha'}{\varepsilon}$ exactly as anticipated above.
The relevance of this scale, instead of $\ell_s$, is a sign of UV/IR mixing as pointed out in the context
of NCSYM theories in \cite{Barbon:2008ut}.

We can now ask about the more precise transition implied by \eqref{mainab}. 
First, we notice that eq.\ \eqref{mainab} can be 
recast (up to an overall numerical factor for each line on the r.h.s.) as 
\beq
\label{finalaa}
\SS_{A_S}(\ell) \sim \Bigg\{
\begin{array}{c}
N^2 \frac{L^2 \ell_c \ell}{\varepsilon^4} +\ldots~~, ~~\ell \ll \ell_c
\\
~\\
N^2 \frac{L^2 \ell_c^2}{\varepsilon^4} +\ldots~~, ~~\ell \gg \ell_c
\end{array} 
~.
\eeq
Moreover, in this language the condition \eqref{mainaaa} becomes simply $\varepsilon\ll \ell_c$.
In a weakly coupled open string theory with a UV cutoff $\varepsilon^{-1}$ and $U(N)$ Chan-Paton indices one expects
the following behaviour of the EE. When $\ell\ll \ell_c$, all the degrees of freedom in the region $A_S$ can interact (and
entangle) with the outside degrees of freedom. In a unit cell of volume $\varepsilon^3$ there are roughly 
$N^2 \frac{\ell_c}{\varepsilon}$ degrees of freedom, so one expects the EE to scale as
\beq
\label{finalab}
\SS_{A_S} \sim \ell L^2 \cdot N^2 \frac{\ell_c}{\varepsilon} \cdot \frac{1}{\varepsilon^3} = N^2 \frac{\ell_c \ell}{\varepsilon^4}
\eeq
reproducing the volume law in the first line of the r.h.s.\ of eq.\ \eqref{finalaa}.
Similarly, when $\ell\gg \ell_c$ only degrees of freedom inside a strip of size $\ell_c$ around the boundary $\p A_S$ can at most entangle with the outside, which leads to the area scaling of the second line of the r.h.s.\ of eq.\ \eqref{finalaa}.
Exactly the same type of scaling and transition (with $\alpha'$ replaced by the non-commutativity scale $\theta$) was argued in the NCSYM theory \cite{Barbon:2008ut,Karczmarek:2013xxa}. Of course, there are also important
differences with the NCSYM theory, which exhibits anisotropy in certain spacetime directions.

The above interpretation suggests that the transition \eqref{finalaa}, from a volume dependence to an area dependence,
is a feature independent of the geometry of the region $A_S$. In agreement with this expectation, we have verified for the cylinder geometry \eqref{disc} that the HEE exhibits the leading divergent terms
\beq
\label{finalac}
\SS_{A_S}(\ell) \sim \Bigg\{
\begin{array}{c}
N^2 \frac{L \ell_c \ell^2}{\varepsilon^4} +\ldots~~, ~~\ell \ll \ell_c
\\
~\\
N^2 \frac{L \ell_c^2 \ell}{\varepsilon^4} +\ldots~~, ~~\ell \gg \ell_c
\end{array} 
\eeq
with a critical length $\ell_c$ again of the order $\sqrt{\lambda} \frac{\alpha'}{\varepsilon}$.

It would be interesting to reproduce the above expectations with an explicit computation in weakly coupled open string theory, perhaps along similar lines to the computation performed in \cite{He:2014gva}. A related computation worth exploring has to do with the corrections of the entanglement entropy of $\NN=4$ SYM theory in the presence of irrelevant deformations induced by open string theory. In the bulk this involves a computation in a regime where the radial cutoff $r_\varepsilon$ is comparable to $r_H$. In this paper we focused on the regime $r_\varepsilon \gg r_H$.

Finally, note that in the computations of this paper the volume law came essentially from the effects of the asymptotic flat space ---the scaling of eq.\ \eqref{gram} was derived exclusively in flat space. We would like to emphasize two points related to this feature that are conceptually close to the discussion of holography as a tomographic principle in \cite{Niarchos:2015moa}. First, we note that although flat space itself does not have a preferred radial direction, the 3-brane solution naturally defines one via the splitting $ds^2=\eta_{\mu\nu}dx^\mu dx^\nu + dr^2 + r^2 d\Omega_5^2$ that leads to the minimal co-dimension-2 surface we considered. For a $p$-brane solution of different worldvolume dimension a different transverse space would be chosen leading to another minimal surface. Second, notice that we would not have obtained sensible results had we restricted only to the flat space part of our computation. In flat space the minimal surface would not exhibit the bottom branch of Fig.\ \ref{ulFigs} and above some width $\ell_{max}$ no minimal surface would exist. In the 3-brane setup this potential issue is remedied by the non-trivial geometry in the bulk that deviates from flat space.

\vspace{0.3cm}
\begin{acknowledgments}
I am grateful to Francesco Aprile, Aristos Donos, Simon Ross and Tadashi Takayanagi for useful comments.
\end{acknowledgments}


\bibliographystyle{utphys}
\bibliography{openclosed}
\end{document}